\documentclass[11pt]{iopart}
\usepackage{graphicx}
\usepackage{epstopdf}
\usepackage{dcolumn}
\usepackage{amssymb}
\usepackage{bm}
\def\spose#1{\hbox to 0pt{#1\hss}}
\def\Vec#1{\mbox{\boldmath$#1$\unboldmath}}
\def\lta{\mathrel{\spose{\lower 3pt\hbox{$\mathchar"218$}}
     \raise 2.0pt\hbox{$\mathchar"13C$}}}
\def\gta{\mathrel{\spose{\lower 3pt\hbox{$\mathchar"218$}}
     \raise 2.0pt\hbox{$\mathchar"13E$}}}
\newcommand{\be}{\begin{equation}}
\newcommand{\ee}{\end{equation}}
\newcommand{\bea}{\begin{eqnarray}}
\newcommand{\eea}{\end{eqnarray}}
\newcommand{\ex}{\mbox{e}}
\newcommand{\dd}{\mbox{d}}

\def\setR{\mathbb{R}}
\def\setC{\mathbb{C}}

\newcommand{\ie}{\textsl{i.e.~}}

\newcommand{\prd}{{\it Phys. Rev.} {\bf D}}
\newcommand{\prl}{{\it Phys. Rev. Lett} }

\begin{document}

\title[K-Bounce]{K-Bounce}

\author{L. Raul Abramo} 
\address{Instituto de F\'{\i}sica, Universidade de S\~ao Paulo,\\
CP 66318, CEP 05315-970, S\~ao Paulo, Brazil}
\ead{abramo@fma.if.usp.br}

\author{Patrick Peter} 
\address{${\cal G}\setR\varepsilon\setC{\cal O}$ --
Institut d'Astrophysique de
 Paris, UMR7095 CNRS, Universit\'e Pierre \& Marie Curie, 98 bis
 boulevard Arago, 75014 Paris, France}
\ead{peter@iap.fr}


\begin{abstract}

By demanding that a bounce is nonsingular and that perturbations are
well-behaved at all times, we narrow the scope of possible models with
one degree of freedom that can describe a bounce in the absence of
spatial curvature. We compute the general properties of the transfer
matrix of perturbations through the bounce, and show that spectral
distortions
of the Bardeen potential $\Phi$
are generically produced only for the small
wavelengths,
although the spectrum of long wavelength 
curvature perturbations 
produced in a contracting phase gets propagated
unaffected through such a bounce.

\end{abstract}

\maketitle

\section{Introduction}

It has become generally admitted, especially with the recent WMAP
data~\cite{WMAP}, that the Universe must have undergone a phase of
inflation~\cite{inflation}, i.e. a very short period of time during
which the ongoing expansion was exponentially accelerated. This phase
not only solves the usual cosmological flatness, homogeneity, monopole
excess and horizon problems, but it also produces, as a bonus, an
almost scale-invariant (usually, but not always, slightly red)
spectrum of primordial scalar perturbations. These small
inhomogeneities, of one part in about $10^5$, have the right spectrum
and can be given, with some amount of fine-tuning, the right amplitude
to seed the large scale structure formation. In fact, it can be
argued~\cite{slava} that the inflationary expansion and the ensuing
superadiabatic amplification of the zero-point energy of the quantum
fields is the only plausible mechanism to transfer microscopic quantum
fluctuations up to the cosmologically relevant length scales without
annihilating the amplitudes of those fluctuations in an ever expanding
universe.

Bouncing models~\cite{bounce,PBB} have been proposed as alternatives to
this scenario, mostly in the framework of string
theory~\cite{ekp,bounce_string} (see, however,
Ref.~\cite{no_ekp}). A bounce, \ie a period of contraction followed by
expansion, could explain the flatness if the expansion phase lasted
much less than the contracting era; it could explain homogeneity by
making the past light cone very large during the contracting era so
thermalization could take place; and, as it turns out, it can very
easily give rise to the same mechanism of superadiabatic amplification
as inflation~\cite{QC_bounce}.

The main distinction of bouncing models compared to inflation lies in
which term dominates the spacetime curvature $R=-12H^2-6\dot{H}$.
Whereas in inflation $\dot{H} \ll H^2$ and the physical wavelengths
grow much faster than the curvature radius $|R|^{-1/2} \sim H^{-1}
\sim \rho^{-1/2}$, close to a bounce $H\simeq 0$ and the curvature radius
grows to $|R|^{-1/2} \sim \dot{H}^{-1/2} \sim [-(\rho + p)]^{-1/2}$ as
the contraction rate grinds to a halt, then falls back down rapidly as
the expansion phase begins.

This means that the modes of interest are pushed inside the curvature
radius during the bounce, and then out again as the universe expands.
This ``in-out" transition is what makes superadiabatic amplification
possible, both in inflation as well as in bouncing models.  The
question is whether sufficiently natural models can be found which
give rise to near-scale invariant spectra of cosmological
perturbations~\cite{QC_bounce}.

As opposed to inflation, in which the phenomenological consequences of
the simplest single-field, slow-roll models are extremely similar
(slightly red spectra), in bouncing models the ensuing spectrum of
cosmological perturbations can vary dramatically, depending on the
model. Moreover, making the universe bounce is far from
straightforward since general relativity forbids this behavior as long
as the Null Energy Condition (NEC) holds. As a result, bouncing models
can become rather intricate.  The simplest bouncing models developed
so far have relied on a combination of fluids~\cite{2fluids},
the presence of spatial curvature~\cite{courb1} or ghost fields 
\cite{courb2}. Some models predict
mode mixing with or without spectral modifications through the bounce
itself, and it has been suggested that these features essentially
originate from either the mixture of two fluids, i.e. from the entropy
perturbations (see ~\cite{2fluids} 
and, in particular, \cite{BozzaVeneziano} where
the precise treatment of entropy perturbations is done), 
or from spatial curvature~\cite{courb1}. 
Hence the need for a single
field flat space bouncing model.

Therefore, we propose here a minimalistic model for a bounce with a
single matter component (a generalized scalar field, or K-essence) and
zero spatial curvature. The condition that spatial curvature is small
shortly after the bounce is a natural one if the contracting phase
lasted much longer than the ongoing expansion phase. We also demand
that the energy density is positive at all times, that a non-singular
bounce takes place, and that the sound speed of perturbations is
well-behaved at all times.

By expressing the Lagrangian of the generalized scalar field as a
Taylor series around the field and its momenta, we can easily
implement these constraints and proceed to construct a very general
class of sensible bouncing models with a single fluid and no spatial
curvature. The only shortcoming of our class of models is that,
because $\dot{H} \geq 0$ near the bounce, they all lie in the
``phantom" sector, $w\equiv p/\rho \leq -1$, so the connection with an
expanding radiation era would necessitate the introduction of matter
fields and a decay mechanism similar to
preheating~\cite{preheat}. This is precisely the scenario recently
proposed in Ref.~\cite{Creminelli}, where an explicit scenario is
realized using the ghost condensate model~\cite{GhostCond}.  For
additional context on the use of non-canonical scalar fields in
cosmology, see, e.g., 
Ref.~\cite{Sen,ST,FT02}. 
Note also that if one
assumes a contracting phase dominated by normal matter (preferably
pressureless matter, in order to get an almost scale invariant
spectrum of perturbations~\cite{ns1_antes}), then because the phantom
divide cannot so easily be crossed~\cite{RaulNelson}, there must also
exist a transition between this contraction and our K-bounce,
equivalent to preheating but in the other way, that one could
henceforth call precooling.

The advantage of our models lies in the simplicity of their
perturbative sector.  We show explicitly that cosmological
perturbations can be propagated in a non-singular way through the
bounce. We also show that the perturbations are well-behaved through
the numerous instantaneous de Sitter phases (moments of time at
which $\dot{H} = 0$) that take place in our model.

We have computed the transfer function for perturbations, and we show
that an initial spectrum of cosmological perturbations can get
distorted by the bounce. As this distortion depends on the duration of
the bounce, our conclusion is that bouncing models generate power
spectra with a wide variety of scale dependences. However, the scale
dependence of the transfer matrix is important only for short
wavelengths, so that the cosmologically relevant (large) scales
are transferred through the bounce unaffected.

This paper is organized as follows. First, we describe the K-essence
model generating the K-bounce, provide the relevant equations of
motion for the background and derive the conditions under which a
bounce is possible (Sec. II). We then specify, in Sec. III, through a
Taylor expansion around the bounce, the form of the pressure function
we use afterwards, and provide the constraints for a 
non-singular bounce to take place. Sec. IV discusses a number of
specific background models and attempts at classifying them by the
bounce duration. We then move on, starting in Sec. V, to the study of
the perturbations. We first reduce the overall system to a single
equation for the only degree of freedom avalaible, which we chose to be 
the Bardeen gravitational potential $\Phi$. We discuss analytically
several potentially problematic cases (the bounce itself, the
instantaneous de Sitter phase, and the quasi de Sitter bounce), and
we show that $\Phi$ is well behaved at all times. Having shown
the propagation of linear perturbations across the bounce to be
regular at all times, we then compute numerically this time evolution,
setting initial conditions at an arbitrary time at which we impose
$\Phi_k=1$ and $\dot\Phi_k=0$ for the Fourier modes. 
We end up with some
considerations about model building in a concluding section.

\section{Generalized scalar-field models}

We will assume that the matter sector is represented by
a scalar field Lagrangian of the form
\be
\label{Lag}
\mathcal{L} = \sqrt{-g} \; p(X,\phi) \; ,
\ee
where
\be
\label{def:X}
X \equiv \frac12 \; g^{\mu\nu} \partial_\mu \phi \; \partial_\nu \phi
\; , \ee and use the timelike signature, ${\rm
diag}(+,-,-,-)$ for the metric $g$. From this Lagrangian, one gets a
stress-energy tensor reading
\be
T^{\mu\nu} = \left(\rho + p\right) u^\mu u^\nu - p g^{\mu\nu},
\label{Tmunu}
\ee
with energy density $\rho = 2 X p_{,X} -p$ and $u_\mu = \phi_{,\mu}
/\sqrt{2X}$. These relations give back the usual one for the canonical
scalar field theory provided one then takes the simplest Lagrangian
function $\mathcal{L}_\mathrm{c}=X-V(\phi)$.

Since Ostrogradski's theorem~\cite{Ostrogradski,Arnold,Woodard}
precludes local higher derivative terms from appearing in the action
principle, Eq. (\ref{Lag}) is the most generic scalar field Lagrangian
which may be stable. Notice that our Lagrangian does not need to be
separable in terms of functions of the kinetic term $X$ and the field
$\phi$, as is sometimes assumed for K-inflation~\cite{garrmukha} or
K-essence~\cite{kessence}. 

The important aspect of the quantum instability of this theory would
also need to be addressed, since evidently any Hamiltonian which is
unbounded from below would be instantly destroyed by quantum tunelling
of positive-energy particles into the negative-energy
particles~\cite{Woodard}. For theories with non-canonical kinetic
terms the quantum stability is a nontrivial issue, in particular for
the case of ``phantom" models -- see, for instance,
Ref.~\cite{RaulNelson}.

Introducing the flat Friedman-Lema\^{\i}tre-Robertson-Walker metric
with scale factor $a(t)$, namely $$
\dd s^2 = \dd t^2 - a^2(t) \dd \Vec{x}^2,
$$ and the Hubble parameter $H\equiv\dot{a}/a$ (a dot standing for a
derivative w.r.t. the time coordinate $t$), the Einstein field
equations are then given by
\bea
\label{EFE0}
3 H^2 &=& 8 \pi G \, \rho = 8 \pi G \,
\left( 2 X p_{,X} - p \right) \; , \\
\label{EFE1}
- 3 H^2 - 2 \dot{H} &=& 8 \pi G \, p \; .
\eea
As for the matter field, the Euler-Lagrange equation stemming from
Lagrangian (\ref{Lag}) is nothing but the conservation of the
stress-energy tensor (\ref{Tmunu}), \ie
\be
\label{EoM0}
\nabla_\mu T^{\mu\nu} = 0 \quad \Longrightarrow \quad p_{,X} \Box \phi +
\partial^\mu \, \partial_\mu \, (p_{,X}) - p_{,\phi} = 0\; , 
\ee
which, under the assumption of homogeneity of the scalar field, $\phi
\rightarrow \phi(t)$, is reduced to the simpler form,
\be
\label{EoM}
\frac{\ddot{\phi}}{c_X^2}
+ 3 H \dot{\phi} 
+ \frac{\rho_{,\phi}}{p_{,X}} = 0 \; ,
\ee
which reduces to the Klein-Gordon equation for the canonical theory
$\mathcal{L}_\mathrm{c}$. Here the sound speed is given by
\be
\label{cs2}
c_X^2 = \frac{p_{, X}}{\rho_{,X}} = \frac{p_{,X}}{2 X p_{,XX}+p_{,X}} \; ,
\ee
and it should be clear from the unapproximated equation of motion that
it is the function responsible for the speed with which inhomogeneous
scalar field fluctuations propagate through spacetime. In particular,
a negative $c_X^2$ would give rise to exponentially growing
small-scale fluctuations, meaning that the theory is classically
unstable.

Since our final aim concerns the predictability and spectrum of
cosmological perturbations before and after the bounce in a one-fluid
model, our first requirement is that the sound speed never becomes
negative.  We also demand that it remains finite, since a diverging
sound speed would cause a singularity in the transfer matrix
\cite{courb1,courb2} that relates the 
cosmological perturbations before and
after the bounce -- destroying, once again, the predictability of the
theory. Therefore, our first physical constraint is
\be
\label{Constr_cs2}
0 \leq c_X^2 < \infty \; .
\ee
Notice that even though we demand that the sound speed 
squared is always positive and finite, we should still work 
under the assumption that our models are just 
phenomenological realizations of some unknown fundamental 
theory, so that the second-quantized perturbations
of the gravitational degrees of freedom are not being taken into 
account properly here. Otherwise, since
both $p_{, X}$ and $2 X p_{,XX}+p_{,X}$ are negative through the bounce
phase in our models, the theory can become unstable, decaying 
instantaneously through graviton production~\cite{Woodard,bruneton}.

Our second requirement is that the energy is non-negative. In 
particular, if the bounce happens at $t=t_0$ we must have that
\be
\label{Rho_B_0}
\rho(t_0)=0 \, ,
\ee
and, as a result of the Einstein equations written in the form
$-3H^2-2\dot{H} = 8 \pi G p$, we conclude that we must impose
\be
\label{Press<0}
p(t_0) = p_0 < 0 \, .
\ee
This means that, as $t \rightarrow t_0$, the equation of state
parameter $w\equiv p/\rho$ becomes infinitely
negative.  But $w<-1$ is the domain of the so-called ``phantom" (or
``ghost") models, and it has been shown that crossing the ``$\Lambda$
barrier" $w=-1$ is impossible in simple single-field models
\cite{RaulNelson,LambdaBarrier}. Therefore, our bouncing model is
limited to $w\leq -1$, which in practice means that in the asymptotic
past (future) the Universe approaches a contracting (expanding) de
Sitter stage. These limiting stages must somehow be connected with
non-phantom dominated epochs through precooling and preheating phases.

\section{Taylor expansion of the Lagrangian}

Our model relies on a series expansion of the Lagrangian in terms of
the field and its momentum. Without loss of generality we set the
value of the field at the bounce to be $\phi(t_0)=0$, and its time
derivative $X(t_0)=X_0\neq 0$, so we can write
\bea
\label{p_exp}
p(X,\phi) &=& p_0 + p_X \, (X-X_0) + p_\phi \, \phi + p_{X\phi} \,
\phi (X-X_0)
\\ \nonumber
& & +  \frac12 p_{XX} \, (X-X_0)^2 
+\frac12 p_{\phi\phi} \, \phi^2 
+ \cdots
\eea
In order to obtain a well-behaved bounce, it is helpful to assume that
the behavior of the field near the bounce is analytic in time
\be
\label{phi_t}
\phi(t) \approx \phi_1 (t-t_0) + \phi_2 (t-t_0)^2 + \phi_3 (t-t_0)^3 + \cdots \; ,
\ee
which means that $X_0=\frac12 \phi_1^2$. We stress that the Taylor
expansion in Eq. (\ref{phi_t}) is not used in any way to constrain the
dynamics -- we only use it as a means to adjust the parameters of the
Lagrangian in light of the constraints.

The constraint that $\rho(t_0)=0$ translates into
\be
\label{constr_1}
2 p_X X_0 - p_0 = 0 \quad \Longrightarrow \quad p_0 = p_X \phi_1^2\; .
\ee
A second constraint comes from the stress tensor conservation, \ie
$\dot\rho = -3 H (\rho+p)$, imposing that $\dot\rho \rightarrow 0$ at
the bounce. This means that only the term in $\rho$ which is quadratic 
in time survives. In terms of our parameters, this condition is expressed as
\be
\label{constr_2}
p_\phi = 2 \phi_2 p_X + \phi_1^2 (p_{X\phi} + 2 \phi_2 p_{XX}) \; .
\ee
Finally, the Friedman equation $8 \pi G p=-3H^2-2\dot{H}$ at $t=t_0$ leads to
the third constraint, namely that $p_X<0$ and $p_0 = p_X \phi_1^2 <0$,
with $p_X$ given by
\be
\label{constr_3}
p_X = \frac{2}{\phi_1^3} \left\{ \phi_3
\pm |\phi_3| \, \left[ 1 - 
p_{\phi\phi} \frac{\phi_1^4}{6\phi_3^2}
+ p_{X \phi} \frac{2 \phi_1^4 \phi_2}{3 \phi_3^2} 
+ p_{X X} \left( 2 \frac{\phi_1^4 \phi_2^2}{\phi_3^2} 
+ \frac{\phi_1^5}{\phi_3} \right) \right]^{1/2} \right\} \; .
\ee
Notice that $\phi_1$
and $\phi_3$ must be chosen such that the square root is real, and such 
that $p_X$ is negative.
Notice also that the $+$ and $-$ branches are identified by simultaneously
changing the signs of $\phi_1$ and $\phi_3$. 

To summarize: our set of constraints determines some relationships
between the Lagrangian parameters $p_0$, $p_\phi$ and $p_X$ in the
context of the class of models in which the behavior of the scalar
field near the bounce can be represented as Taylor series. Presumably,
other models for which the scalar field around the bounce cannot be
represented by such a series will lead to similar constraints between
the parameters involved in these cases. As we are interested in some
minimalistic bouncing model and the general conclusions that can be
drawn thereof, this will suffice for us.

Notice that the scalar field parameters $\phi_i$ are essentially free,
and that the Lagrangian parameters $p_{\phi\phi}$, $p_{X\phi}$ and
$p_{XX}$ are also essentially free -- we only need to make sure that
the square root in Eq. (\ref{constr_3}) remains real. Higher-order
parameters ($p_{XXX}$ etc.) would come into these constraints, but
they would also remain basically free.  This means we can tune the
parameters of the Lagrangian in order to make the models stable --
which is very important for phantom models.  It also means that we can
set the parameters such that the bounce is short or long, fast or
slow, at will.

\section{Concrete models: background}

The class of models one can construct with the procedure above has a
very rich phenomenology. In all of them the conditions we impose on
the parameters are such that any bounce, defined as the point in time
at which $H=0$, is necessarily non-singular, having $\dot{H}>0$ at
this point. Therefore, if we set our initial conditions to a Universe
that is contracting, it necessarily will end up bouncing provided the
constraints on the underlying parameters are indeed satisfied. 

There is only one kind of fixed point in our theory, namely 
$\dot{H}\rightarrow 0$. As a
result, and whatever the initial conditions, once we have passed
through the bounce, our models necessarily asymptote to a de Sitter
Universe; this fixed point is an attractor provided $H>0$, and a
repulsor otherwise. In practice, some intermediate quasi-de Sitter
phases (contracting as well as expanding) can happen as the model
contracts and then expands, which is rather interesting from the point
of view of the background model, but represents a formidable
complicating task if one is interested in the perturbations.

We have chosen to concentrate on three concrete models, one in which
the bounce is relatively fast and short, one in which it is a slow and
long phase, and another in which we tuned the parameters so that the
bounce is also a quasi-de Sitter phase (i.e., both $H=0$ and
$\dot{H}=0$ at the bounce.)  All models approach a contracting
(expanding) de Sitter phase in the past (future), which is natural
since going backwards in time transforms the repulsor with $H<0$ into
an attractor. The contracting phase is in fact an unstable point which
all trajectories exit from, whereas the expanding de Sitter phase is
an attractor point where all our models must finish. Therefore, in
order to make the transition to a radiation-dominated Universe we must
introduce new ingredients, or make the scalar field decay into some
other fields. As this reheating process usually preserves the basic
properties of the cosmological perturbations (at least in the single
field case at hand), we will not treat it here -- see, for instance,
Ref. \cite{Creminelli}. Similarly, if we want to originate with a
stable-matter dominated phase, we will need a transition ({\it
precooling}) to lead into the bounce phase. For the same reasons as
the preheating, we shall not consider the details of such a
transition, and will just assume, as usual, that the long-wavelength
spectrum of perturbations is transmitted unchanged through these
precooling and preheating phases. Technically, this translates into
saying that we do not impose physically motivated initial conditions
here, assuming that they have been generated in the phase preceeding
the precooling, and therefore out of the scope of this paper.

It is useful to write down a few identities for the background
that hold in general. The equation of state can be written, with
the help of Friedmann equations, as
\be
\label{eqstate}
w=\frac{p}{\rho} = -1 -\frac23 \frac{\dot{H}}{H^2} \; ,
\ee
while its time derivative can be conveniently expressed as
\be
\label{dotw}
\frac{\dot{w}}{1+w} = \frac{\ddot{H}}{\dot{H}} - 2 \frac{\dot{H}}{H} \; ,
\ee
where we have used the continuity equation, $\dot\rho = -3 H \rho (1+w)$.


\section{Perturbations}

We now perturb the scalar field as $\phi \rightarrow \phi(t) +
\delta\phi(\Vec{x},t)$, and for the metric we fix the gauge to the 
conformal-newtonian (longitudinal) one as~\cite{MFB}
\be
\label{long}
\dd s^2 = \left[ 1+ 2 \Phi \left( \Vec{x},t\right) \right] \dd t^2 -
\left[ 1-2\Phi\left( \Vec{x},t\right) \right] a^2(t) \dd\Vec{x}^2 \; .
\ee
By using the constraint equations in the case of a single generalized
scalar field we can express the Mukhanov-Sasaki variable
\cite{garrmukha} as
\be
\label{v}
v = z \, \zeta \; ,
\ee
where
\be
\label{zeta}
\zeta \equiv \Phi + H \frac{\delta\phi}{\dot\phi} =  \Phi - \frac{H}{\dot{H}} 
(\dot\Phi+H\Phi) \; ,
\ee
and in our case ($w < -1$) we have
\be
\label{z}
z^2\equiv - \frac32 \frac{a^2 (1+w)}{c_X^2} \; .
\ee

The Mukhanov variable $v$ obeys the equation
\be
\label{eq:v}
v'' + \left( c_X^2 k^2 - \frac{z''}{z} \right)v = 0 \; ,
\ee
where a prime denotes a derivative with respect to conformal time
$\eta = \int \dd t/a$, \ie $\dd/\dd\eta=a \dd/\dd t$. In lieu of
Eq.~(\ref{eq:v}) we can view $z''/z$ as an effective potential that is
scattered by the incoming wave $v$.

However, it can be immediately seen from Eq. (\ref{z}) that the
transformation to the variable $v$ is ill-defined in two particularly
important situations: first, if the equation of state goes to
infinity, as happens in our bounce, and second, if $w\rightarrow -1$,
as happens if the Universe reaches a de Sitter phase. 
In fact, the effective potential $z''/z$ becomes singular
in these situations. Obviously, in these cases the Mukhanov variable
cannot be usefully employed and we must search for other, more
suitable ways to represent the perturbations. It is interesting to
realize that the situation is similar to what happens in the curvature
dominated bounce examined in Ref.~\cite{courb1}: the Mukhanov variable
becomes useless in both these bounce cases. We must therefore resort
to the original Einstein equations for the metric perturbations
directly.

In the single scalar field case it is possible to write a second-order
differential equation for $\Phi$, which reads
\be
\label{eq:Phi}
\ddot\Phi 
+ \left( H - \frac{\ddot{H}}{\dot{H}} \right) \dot\Phi 
+ \left(c_X^2 \frac{k^2}{a^2} + 2\dot{H} - 
 \frac{H\ddot{H}}{\dot{H}} \right) \Phi 
= 0 \; .
\ee
It is clear that this equation is completely well-behaved
through a bounce ($H \rightarrow 0$), as long as $\dot{H}$ remains
finite. 

As is well known, this equation also describes well the perturbations
in a nearly de Sitter (inflationary) spacetime.  This is evident if we
write Eq. (\ref{eq:Phi}) in terms of the slow-roll parameters
$\epsilon \equiv - \dot{H}/H^2$ and $\delta \equiv -
\ddot{H}/(2H\dot{H})$
\be
\label{eq:Phi_SR}
\ddot\Phi 
+ \left( 1 + 2\delta \right) H \dot\Phi 
+ \left(c_X^2 \frac{k^2}{a^2 H^2} - 2\epsilon + 2\delta \right) H^2 \Phi 
= 0 \; .
\ee
It then becomes obvious that by taking $\epsilon \rightarrow 0$ and
$\delta \rightarrow 0$ and neglecting the exponentially decaying
gradient term we obtain that in the slow-roll regime
\be
\label{Phi_SR}
\Phi_{\rm S.R.} \sim A \ex^{- (1-\delta+2\epsilon) H t} + 
B \ex^{-2 (\delta-\epsilon) Ht} \; ,
\ee 
to first order in the slow-roll
parameters.

However, Eq. (\ref{eq:Phi}) may not be appropriate 
in an instantaneous de Sitter point, i.e., an instant of time
when $w = -1$ ($\dot{H} = 0$.).
We have found that it is particularly enlightening to write
the following set of first-order equations
\bea
\label{seteq1}
\frac{\dot{H}}{H} \dot\zeta &=& c_X^2 \frac{k^2}{a^2} \Phi \; ,
\\
\label{seteq2}
\dot\Phi + H \Phi &=& - \frac{\dot{H}}{H} (\zeta-\Phi) \; .
\eea
These relations, together with Eq. (\ref{zeta}), 
show that the perturbations are propagated
through the many important phases described below in a regular way.

\subsection{Exact solution near the bounce}

First, consider a bounce at $t=0$ that occurs within the class of
models given by Eqs. (\ref{p_exp})-(\ref{phi_t})\footnote{Here and in
the following subsection, the choice $t=0$ for the point under
consideration is of course a mere convention aimed at simplifying the
subsequent equations; in the numerical approach, we will set the
initial contracting solution at $t=0$, so the bounce takes place at a
different location.}. We can then write
\be
\label{H_bounce}
H\approx H_1 t + \frac12 H_2 t^2 + \frac16 H_3 t^3 + \cdots \; ,
\ee
where we take $H_1>0$ in accordance with our class of models
for which $1+w \leq 0$.
We suppose the approximation above to be valid for small times such that
$|t| \ll \min \left( 2 H_1/|H_2|, 3 |H_3|/|H_2| \right)$. 

Substituting the approximation above into Eqs. (\ref{seteq1})-(\ref{seteq2}) 
or, equivalently, into Eq. (\ref{eq:Phi}), and keeping only
the dominant terms we obtain the following equation for $\Phi$:
\be
\ddot\Phi + H_1 \left[ \left( 1 + \frac{H_2^2}{H_1^3} 
- \frac{H_3}{H_1^2} \right) \, t - \frac{H_2}{H_1^2} \right] \, \dot\Phi 
+ \left( c_X^2 \frac{k^2}{a_0^2} + 2 H_1 + H_2 t \right) \Phi \approx 0\; ,
\label{eqPhiBounce}
\ee
where $a_0$ is the (near-constant) scale factor at the bounce,
so we implicitly assume that $|t| \ll H_1^{-1/2}$. We will
also assume that the sound speed $c_X^2$ is approximately constant
during the bounce, which is the case in all models we have considered.
It is interesting to notice that if $H_2=0$, then there is a limiting case 
$H_3=H_1^2$ ($\ddot{H}=\dot{H} H$) for which the exact solutions near 
the bounce are pure oscillatory modes.

By writing the solution to Eq. (\ref{eqPhiBounce}) as a truncated
Taylor series in time, it is easy to find two linearly independent
approximate solutions,
\bea
\label{Bapprox}
\Phi_1 &\approx& 
1 
- \left( 1 + \frac12 \gamma_k^2 \right) H_1 t^2 
- \frac{H_2}{6}\left( 3 + \gamma_k^2 \right) t^3 \; ,
\\ 
\nonumber
\Phi_2 &\approx& t + \frac{H_2}{2H_1} t^2
- \frac{H_1}{6} \left( 3 + \gamma_k^2 - \frac{H_3}{H_1^2} 
\right) t^3 \; ,
\eea
where $\gamma_k^2 = c_X^2 k^2/(a_0^2 H_1)$.  We can in fact find exact
solutions to Eq. (\ref{eqPhiBounce}), and the two linearly independent
modes turn out the be essentially a Hermite polynomial and a confluent
hypergeometric function $\Phi$. Both functions are analytic at $t=0$
and reduce, to lowest order in $t$, to the approximate solutions
(\ref{Bapprox}).

In terms of the curvature fluctuation $\zeta$, the approximate
solutions are:
\bea
\label{zeta_B}
\zeta_1 &\approx& 1 + \frac12 \gamma_k^2 H_1 t^2 - \frac{H_2}{6}
\gamma_k^2 t^3 \; , 
\\ \nonumber
\zeta_2 &\approx& \frac13 \gamma_k^2 H_1 t^3 \; . 
\eea
Therefore, the curvature perturbation is completely regular across the
bounce.  Notice that neither the growing nor the decaying modes of the
curvature fluctuations near the bounce depend on the cubic term $H_3$
in Eq. (\ref{H_bounce}), even though the newtonian potential $\Phi$
does.  From Eqs. (\ref{zeta}) and (\ref{Bapprox})-(\ref{zeta_B}) we
can also see that by keeping only the dominant mode we make the
curvature fluctuation $\zeta$ equal to $\Phi$ at the bounce.

Notice that what was the growing mode in the contracting era becomes
the decaying mode in the expanding era, and vice-versa. This behavior
is completely generic for the linear cosmological perturbations, and
has been shown to work in much more complex bouncing models
\cite{GrowDecay}.

\subsection{Exact solution near an instantaneous de Sitter phase}

Now we analyse the solution near a de Sitter point -- i.e., and
instant of time when $w=-1$. 
Since we assume that the sign of $1+w=-2\dot{H}/3H^2$ does not change, we take 
the following approximation for the Hubble parameter near a de Sitter point
which we place at $t=0$:
\be
\label{dSp}
H \approx H_0 + \frac16 H_3 t^3 + \frac{1}{24} H_4 t^4 + \cdots \; ,
\ee
where we suppressed the linear term because we want $\dot{H}=0$ at
$t=0$, and the absence of a quadratic term is implied by $1+w\leq 0$.
The approximation is valid for $|t| \lesssim |6 H_0/H_3|^{1/3}$; we
have found that it is always valid near de Sitter points in our class
of models.

Neglecting the subdominant terms we obtain the following equation for
$\Phi$:
\be
\ddot\Phi +\left( A  - \frac2t \right) \dot\Phi
+ \left(B - \frac{2H_0}{t} \right) \Phi = 0 \; ,
\label{eqPhidS}
\ee
with 
$$
A = H_0 - \frac{H_4}{3 H_3} , \quad \hbox{and} \quad B_k = c_X^2
\frac{k^2}{a^2} - \frac{H_0 H_4}{3 H_3}.
$$
Notice that the cubic term $H_3$ does not show in the equation for
$\Phi$ at leading order order -- it will, however, reappear when we
compute $\zeta$.

Defining
\be
\nonumber
\Omega_k=\sqrt{1- 4 \frac{B_k}{A^2}} \; \; ,
\ee
rescaling the time to $z = A \Omega_k t$, and making the variable
change $\Phi = z^3 \exp{[-(1+\Omega_k) z / (2 \Omega_k)]} y(z)$, we can
reduce Eq. (\ref{eqPhidS}) to the equation for the confluent
hypergeometric function,
\be
\label{eqCHF}
z y'' + (\gamma - z) y' - \alpha y = 0 \; ,
\ee
where $\gamma=4$ and $\alpha=2+(2H_0-A)/(A\Omega_k)$.  The two
linearly independent solutions to Eq. (\ref{eqCHF}) are given by
\bea
\label{CHFs}
y_1 &=& \, _1 F_1 (\alpha,\gamma,z) \; ,
\\ 
\nonumber
y_2 &=& C \, _1F_1(\alpha,\gamma,z) \ln{z} + z^{-3}
\sum_{n=0}^{\infty} v_n z^n\; ,
\eea
where the coefficients $C$ and $v_n$ can be found in standard
textbooks on special functions~\cite{SlaterBook}.  Notice that the
first solution is well-behaved everywhere, but the second solution is
non-analytic at the origin due to the presence of the $\log$ term.
This happens because the confluent hypergeometric
function $_1F_1(\alpha,\gamma,z)$ with integer $\gamma$ has a branch
cut in the Riemann plane. As a result, we could not have found the
{\it second} solution by writing a naive Taylor series around
$t=0$, as was done in the previous section. Nevertheless, specifying
the value of the second function and its derivative anywhere fixes its
value everywhere in the Riemann plane, so the solution can be
propagated from negative to positive values of $t$. 
Of course, Eq. (\ref{eqPhidS}) is
only approximate, so the exact solution to the exact equation may be
much better behaved, but this subtlety rendered the actual numerical
evolution of the perturbative equations tremendously complicated at
de Sitter points.

In terms of the newtonian potential we find the following approximate
solutions for small $z$, namely
\bea
\label{PhidS}
\Phi_1 &\approx& 
\left( 4 - z \right) z^3 + {\cal{O}}(z^5) \; ,
\\ \nonumber
\Phi_2 &\approx& 1- z + c_2 z^2 - c_3 z^3 + 
\left( c_4 + d_4 \ln z \right) z^4 + \cdots \; 
\eea
The first is a decaying (growing) mode in the contracting (expanding)
phase, while the second is a constant mode.

Upon substitution of the modes (\ref{PhidS}) into the definition of
$\zeta$ we find
\bea
\label{zetadS}
\zeta_1 &\approx& a_1 - b_1 t^2 + {\cal{O}}(t^3) \; , \\ \nonumber
\zeta_2 &\approx& t^{-1} [ a_2 - b_2 t^2 + {\cal{O}}(t^3) ] \; .
\eea
The first solution is nothing but the constant curvature mode that
passes essentially unaltered through the instantaneous de Sitter
phase, while the second solution has a pole $\sim t^{-1}$ at the de
Sitter point but has no constant piece.  This pole corresponds to no
real physical singulariy: it just points out the inadequacy of the
definition of the curvature fluctuation in this situation since the
solution for the newtonian potential $\Phi$ is completely well-behaved
and can be propagated through any de Sitter point.

\subsection{Exact solution near a quasi-de Sitter bounce}

An interesting limiting case happens when the bounce occurs
in such a way that, as $H \rightarrow 0$, $\dot{H} \rightarrow 0$
as well. Since $\dot{H} \geq 0$ in our types of models, we conclude
that near this quasi-de Sitter bounce we also have $\ddot{H} \rightarrow 0$.
Therefore, near the quasi-de Sitter bounce the Hubble expansion
parameter can be expanded as
\be
\label{H_qdSB}
H \approx \frac16 H_3 t^3 \; ,
\ee
which, substituted into Eq. (\ref{eq:Phi}), leads to
\be
\label{eqPhi_qdSB}
\ddot{\Phi} - \frac{2}{t} \dot{\Phi} + \beta_k^2 \Phi \approx 0 \; ,
\ee
with $\beta_k^2 = c_X^2 k^2/a_0^2$. There are trivial solutions to this 
equation in terms of spherical Bessel functions $j_{\pm \frac32}$,
\be
\label{solPhi_qdSB}
\Phi \approx
A \tau \left( \frac{\cos\tau}{\tau} + \sin \tau \right)
+ B \tau \left( \cos \tau - \frac{\sin \tau}{\tau} \right) \; ,
\ee
where $\tau = \beta_k t$.
The approximate solutions for the curvature perturbation $\zeta$
around the bounce are then given by
\be
\label{zeta_qdSB}
\zeta \approx A \left( 1 + \frac16 \tau^2\right) 
+ B \left( -\frac{1}{45} \tau^5 \right) \; .
\ee
So, again we see the presence of a constant mode, and of
another mode which decays rapidly in the contracting phase 
but grows rapidly in the expanding phase.

\subsection{Numerical evolution in some concrete models}

We now study a few concrete models. It is important to keep in mind
that the condition that $1+w \leq 0$ (or, equivalently, $\dot{H}>0$)
at the bounce means that the equation of state must be phantomlike at
all times, as crossing the phantom barrier is prohibited in
single-field models \cite{RaulNelson}. Since in the phantom case the
only stable fixed point is $w \rightarrow -1$, it is only natural that
the asymptotic solutions are quasi-de Sitter (contracting and
expanding), as we indeed find.

Nevertheless, our freedom to set the parameters of the Lagrangian
($p_0$, $p_\phi$, $p_X$, etc.) means that we can vary the duration of
the bounce. Since $H=0$ at the bounce, the time scale which sets how
fast the bounce occurs is naturally given by $\Delta T_B \sim
\sqrt{\dot{H_B}}$. Therefore, by tweaking $\dot{H}_B$ we can construct
models in which the bounce is very fast or very slow.

As we are ultimately interested in the cosmological perturbations and
their spectra in these models, it is useful to consider what should
happen to the perturbations if the bounce is very fast or if it is
very slow. Eq. (\ref{eq:v}) tells us that we can regard the problem of
the propagation of cosmological perturbations as that of the
scattering of a wave function $v$ by a potential $V_{v} = - z''/z$. By
changing the duration of the bounce, we are in effect changing the
potential $V_v$ and changing the interval of time in which the wave
function interacts with the potential. 
Hence, intuitively we should
expect that for very fast bounces the short wavelengths will barely reach
reach the oscillatory regime, while for slow bounces
the oscillating stage will be fully realized by the short wavelengths.
Hence, we should expect that, for those wavelengths that can
reach the oscillatory regime, the change in their amplitudes is going
to be more drastic in slow bounce models than in fast bounce models -- see,
later, Fig. 5.

We have constructed three models which are broadly representative
of the phenomenology of K-matter bounces: a fast bounce (FB), a
medium bounce (MB) and a slow bounce (SB).

\begin{figure}
\center
\includegraphics[width=12.5cm]{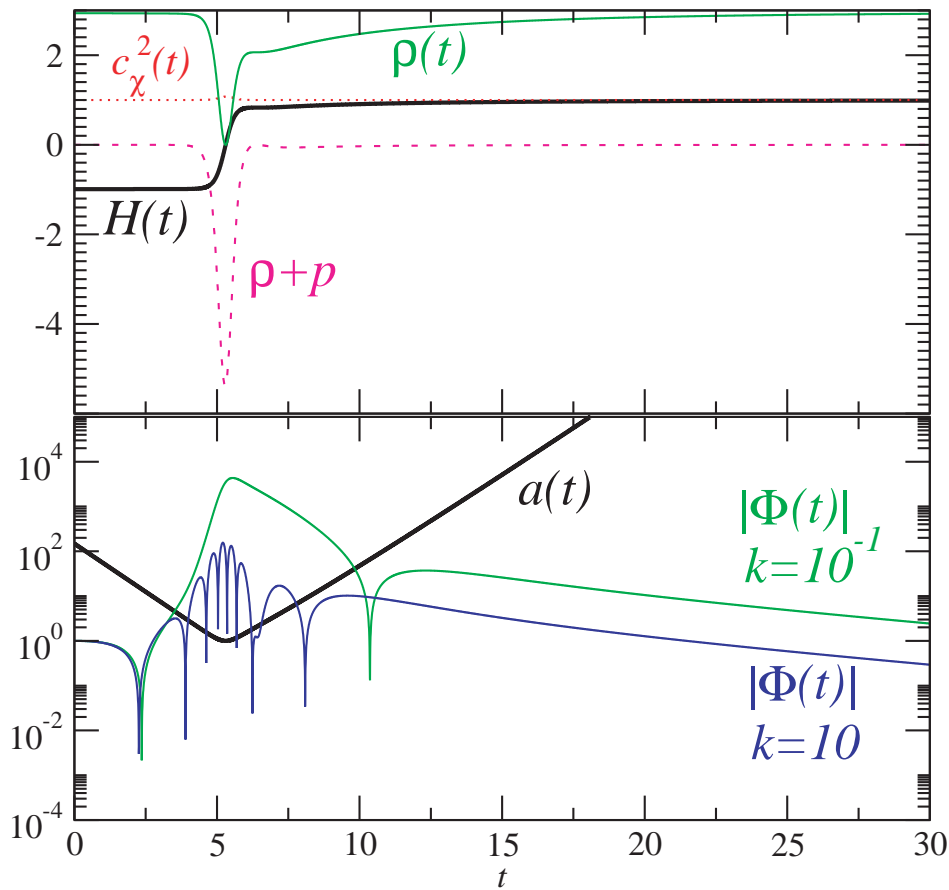}
\caption{\label{fig:1}
Fast bounce model (FB.) In this model the parameters are:
$p_0=-3.815$, $p_\phi=-0.891$, $p_X=-4.710$, $p_{\phi\phi}=1.0$,
$p_{\phi X} =-2.4$, $p_{XX}=0.5$ (with $f_1=0.9$, $f_2=0.1035$ and
$f_3=-1$.)  Upper panel: background quantities $H$, $\rho$, $\phi$,
$c_X^2$ and $\rho+p$.  Lower panel: scale factor $a$ and perturbations
in the metric ($\Phi$) for two different wavelengths, $k_1=10^{-1}$
and $k_2=10$. Notice the kinks in the absolute value of $\Phi_{k_2}$,
which indicate a regime of oscillations. The kinks in $\Phi_{k_1}$,
on the other hand indicate that it starts positive, 
then, during the contracting era
at around $t\sim 2.5$ it becomes negative, and then it becomes 
positive again in the expanding era at $t\sim 11$. The kinks in
$\Phi_{k_1}$ are
a manifestation of the changing roles of the decaying and growing
modes before and after the bounce.}
\end{figure}

First, consider the fast bounce (FB) of Fig. 1. As shown in the upper
panel, the universe starts in a quasi-de Sitter contracting phase,
with $w=-1$ and $H\sim -1.1$.  It contracts with that initial rate up
until $t\sim 4.8$, then it bounces at $t\sim 5$ as the expansion rate
grows very rapidly. It then reaches another quasi-de Sitter phase,
albeit an expanding one.  In our arbitrary time units, the bounce
lasts about $\Delta t \sim 1$.  Notice that at $t\sim 6$ the density
and the expansion rate become flat for an instant of time, meaning
that at that point the equation of state reached the value $w=-1$ 
-- that is, the universe went through an instantaneous de Sitter point.
Notice also that nothing special happens to the sound speed $c_X^2$ --
indeed, in all our models the sound speed is well-behaved and is not
crucial to any of our discussions.

The cosmological perturbations in the FB model are shown in the lower
panel of Fig. 1, for two different wavelengths ($k_1=10^{-1}$ and
$k_2=10$), along with the scale factor. As discussed above, we do not
have a natural criterium to impose on the initial conditions of the Bardeen
potential. Thus we chose, for all
numerically evolved models below, to set $\Phi_\mathrm{ini}=1$ and
$\dot\Phi_\mathrm{ini}=0$:  then, getting anything else but a constant
$\Phi$ for asymptotically long times after the bounce would be evidence 
of mode mixing.

It can be seen that the newtonian potential is well-behaved at all
times (as shown in the analytical solutions of the previous
sections). For wavelengths longer than that of the mode $k_1$ the
solutions for the perturbations are all identical, meaning that the
bounce does not affect them differently.  With our initial conditions
all perturbations go through a sign change at around $t\sim2.5$, which
is just a manifestation of the relative growth of the dominant mode
compared to an initially mixed-mode state.

Notice that the small-wavelength mode $k_2$ detaches from the
behavior of the mode $k_1$ at around $t\sim 5$, which indicates that
the small-wavelength mode almost reaches the oscillatory regime.
Indeed, for wavelengths smaller than that of the mode $k_2$ the
perturbations go through a period of oscillations which becomes longer
as we consider smaller wavelengths. This means that these modes are
small enough to be insensitive to the curvature radius created by the
bounce. Equivalently, we can say that the modes $v$ experience a very
small effective potential $V_v$, so they simply oscillate.

\begin{figure}
\center
\includegraphics[width=12.5cm]{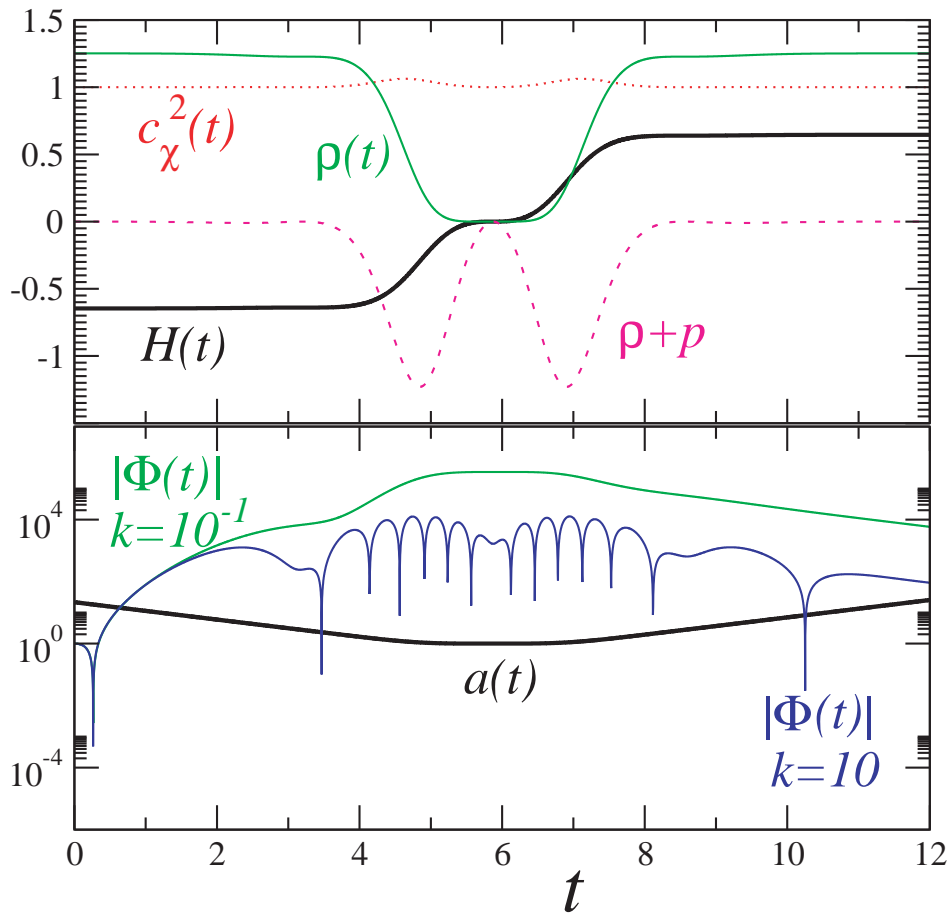}
\caption{\label{fig:2}
Medium bounce model (MB). In this model the parameters are:
$p_0=-1.776$, $p_\phi=-1.915$, $p_X=-1.776$, $p_{\phi\phi}=2.3$,
$p_{\phi X} =0$, $p_{XX}=0.5$ (with $f_1=-1$, $f_2=0.75$ and $f_3=1$.)
Upper panel: background quantities $H$, $\rho$, $\phi$, $c_X^2$ and
$\rho+p$.  Lower panel: scale factor $a$ and perturbations in the
metric ($\Phi$) for two different wavelengths, $k_1=10^{-1}$ and
$k_2=10$.}
\end{figure}

In Fig. 2 we show a bounce model (MB) in which the bounce itself
happens over a longer period of time, $\Delta t \sim 4$. We have also
set the parameters so that the instant of the bounce coincides with an
instant when $\rho+p\rightarrow 0$. Hence, in this model the bounce
($H=0$) is also a quasi-de Sitter point ($\dot{H}=0$); in other words,
the background behaves, close to the bounce, like Minkowski
spacetime. It is interesting, although not entirely unexpected, that even
in this critical model the perturbations are entirely well behaved at
all times.

\begin{figure}
\center
\includegraphics[width=12.5cm]{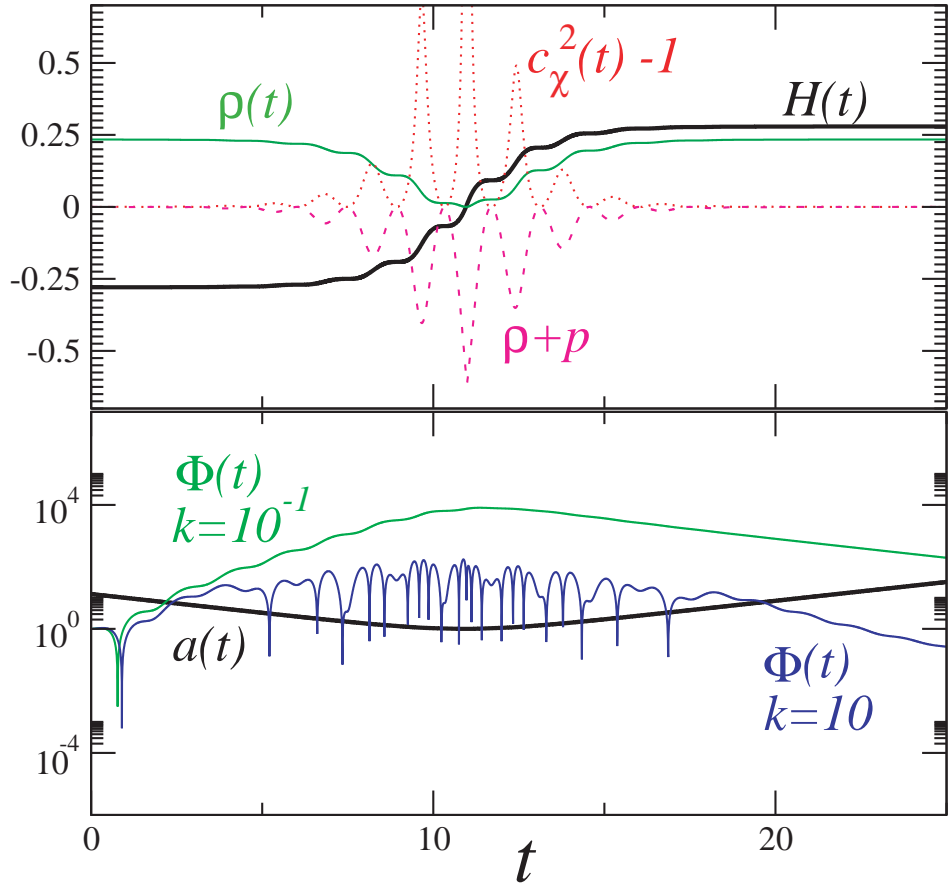}
\caption{\label{fig:3}
Slow bounce model (SB). In this model the parameters are:
$p_0=-0.586$, $p_\phi=-0.172$, $p_X=-0.586$, $p_{\phi\phi}=2.0$,
$p_{\phi X} =0$, $p_{XX}=0.5$ (with $f_1=-1$, $f_2=1$ and $f_3=1$.)
Upper panel: background quantities $H$, $\rho$, $\phi$, $c_X^2$ and
$\rho+p$.  Lower panel: scale factor $a$ and perturbations in the
metric ($\Phi$) for two different
wavelengths, $k_1=10^{-1}$ and $k_2=10$.}
\end{figure}

In Fig. 3 we show the SB model. Here the bounce is accompanied by many
quasi-de Sitter instantaneous points ($\rho+p=0$.)  The bounce happens
during a time scale of $\Delta t \sim 10$.  The most telling
characteristic of the perturbations is that now the mode with
$k_2=10$ experiences more than 20 oscillations during the bounce, 
while in the MB model it only had time to perform 
about six oscillations.

The main result of this Section is that cosmological perturbations pass through 
the bounce with their spectrum essentially unchanged. However, our numerical
evolution cannot address the important question whether there is mixing
between the dominant and sub-dominant modes, before and after the bounce.
This is due to the rapidly decaying nature of the sub-dominant solution
after the bounce.

Let us consider the possibility of mode mixing by means of a
simplified analytical model inspired by the numerically solved fast
bounce scenario.  Let us take the following model for the Hubble
parameter:
\be
H = \frac{H_++H_-}{2} + \frac{H_+-H_-}{2} \tanh \frac{t}{t_0},
\label{Hsimpl}
\ee
which interpolates smoothly between a de Sitter phase with 
contraction rate $H=H_- <0$ and an expanding de Sitter phase with
expansion rate $H=H_+>0$. Notice that we have neglected the term
$\sim k^2$, since the numerical analysis have shown that it only
matters for perturbations of very small wavelength. Eq. (\ref{eq:Phi})
then becomes:
\bea
\ddot\Phi +\left[\frac{H_++H_-}{2} + \left(\frac{2}{t_0}
+ H_+ - H_-\right) \tanh\frac{t}{t_0} \right]
\dot \Phi 
\\ \nonumber
+\left[ H_+-H_- +\left(H_++H_-\right) \tanh\frac{t}{t_0}  
\right] \frac{\Phi}{t_0} = 0.
\eea

In order to get an analytical solution we assume that $H_+=-H_-=h$,
and we take $t_0=1$ for simplicity. (Note that this is strictly
equivalent to introducing a new variable $t/t_0$ and rescalling all
the constants accordingly.)  With these choices, the two linearly
independent solutions to Eq.~(\ref{eq:Phi}) are:
\bea
\label{SolHsimpl_1}
\Phi_1 &=& \left( 1 - z^2 \right)^{(1+h)/2} 
\, P_{h}^{\beta} \left( z \right) \; ,
\\
\label{SolHsimpl_2}
\Phi_2 &=& \left( 1 - z^2 \right)^{(1+h)/2} 
\, Q_{h}^{\beta} \left( z \right) \; ,
\eea
where $z=\tanh t$, $\beta \equiv \sqrt{1+h^2} > 1$, and 
$P_\mu^\nu(z)$ and $Q_\mu^\nu(z)$ are the associated
Legendre functions of the first and second kind,
respectively~\cite{AS}.

Since we would like to connect this universe model with a previous
contracting phase (before pre-cooling) and an ensuing expansion phase
(after pre-heating), we should consider what happens with the two
modes above both at early times ($t \rightarrow -\infty$) and at late
time ($t \rightarrow +\infty$). This is necessary if we give initial
conditions for the perturbations and their time derivatives at some
initial (early) time, and if we would like to follow their evolutions
at late times and ask whether that choice of initial values implies a
mixture of the two modes of
Eqs. (\ref{SolHsimpl_1})-(\ref{SolHsimpl_2}).

It turns out that the asymptotic limits $t \rightarrow -\infty$ ($1+z
\rightarrow 0$) and $t \rightarrow +\infty$ ($1-z \rightarrow 0$) are
very subtle for the associated Legendre functions: in fact, the limit
$|z|\to 1$ is singular in the Legendre differential equation, which
translates into the fact that the Legendre functions are almost
degenerate -- and, of course, without a complete basis of linearly
independent functions we cannot accomodate an arbitrary set of initial
condition.  As a result, we need to expand the functions to
3$^{\rm rd}$ order in the small parameters $(1 \pm z) \sim
\exp(-2h|t|)$ so as to break that degeneracy.  The result is that, in the
limit $t \rightarrow -\infty$ ($1+z \rightarrow 0$), we get:
\bea
\label{Phi_1_minus}
\Phi_1^- &\simeq& 
- \frac{2^{1+\beta/2} \csc \beta\pi \sin h\pi }{\Gamma(1-\beta)}
\\ \nonumber
&\times& 
\left\{ (1+z)^{1-\beta/2}
- \frac{\beta(\beta-1) + 2h}{4}
\frac{\Gamma(1-\beta)}{\Gamma(2-\beta)} (1+z)^{2-\beta/2} \right.
\\ \nonumber
&+& \left. 
\frac{1}{\sin h\pi} 
\frac{2^{1-\beta \pi} \Gamma(1-\beta)}
{\Gamma(-h-\beta)\Gamma(1+h-\beta)\Gamma(1+\beta)}
 (1+z)^{1+\beta/2} + \cdots \right\} \; ,
\\
\label{Phi_2_minus}
\Phi_2^- &\simeq& 
- \frac{2^{\beta/2} \pi \csc \beta\pi \cos h\pi }{\Gamma(1-\beta)}
\\ \nonumber
&\times& \left\{ (1+z)^{1-\beta/2}
- \frac{\beta(\beta-1) + 2h}{4}
\frac{\Gamma(1-\beta)}{\Gamma(2-\beta)} (1+z)^{2-\beta/2} \right.
\\ \nonumber
&+& \left. \left[ 
\frac{1}{\sin h\pi} + \frac{\cot (h+\beta)\pi}{\cos \beta\pi} \right]
\frac{2^{-\beta/2} \pi \Gamma(1-\beta)}
{\Gamma(-h-\beta)\Gamma(1+h-\beta)\Gamma(1+\beta)}
 (1+z)^{1+\beta/2} + \cdots \right\} \; .
\eea
Here we have implicitly assumed that $1 < \beta < 2$, so the
series in non-integer powers of $(1+z)$ is ordered correctly.

Comparing the first two terms in Eqs. 
(\ref{Phi_1_minus})-(\ref{Phi_2_minus}) one can see that they are
identical, up to a constant. Only the third-order terms in these 
solutions have different factors. Therefore, if we want to assign
arbitrary initial conditions to the perturbations at very early
times, we need to go to third order in the series around $(1+z)$.
Since this small factor goes exponentially to zero as
$\exp [4(2-\beta/2)ht_1]$, this means
that numerically it is very hard to select only one of the modes.
Any choice of initial conditions that selected one mode
at the expense of the other, if made at a very early time 
$t_1 \ll -1/h$, would imply a fine-tuning of order 
$\exp [4(2-\beta/2)ht_1]$. This means that quite generically, any 
natural choice of initial conditions at very early times 
will necessarily select a mixture of the two modes, $\Phi_1^-$ 
and $\Phi_2^-$.

Consider now the limit $t \rightarrow +\infty$ 
($1-z \rightarrow 0$):
\bea
\label{Phi_1_plus}
\Phi_1^+ &\simeq& 
- \frac{2^{1+\beta/2}}{\Gamma(1-\beta)}
\times \left\{ (1-z)^{1-\beta/2} \right.
\\ \nonumber
&-& \frac{\beta(\beta-1) + 2h}{4}
\frac{\Gamma(1-\beta)}{\Gamma(2-\beta)} (1-z)^{2-\beta/2}
\\ \nonumber
&-& \left. 
\frac{\beta^3 + 4(1-\beta)- (1+h)^4}{2^5} 
\frac{\Gamma(1-\beta)}{\Gamma(3-\beta)}
 (1-z)^{3-\beta/2} + \cdots \right\} \; ,
\\
\label{Phi_2_plus}
\Phi_2^+ &\simeq& 
\frac{2^{\beta/2} \pi \cot \beta\pi }{\Gamma(1-\beta)}
\times \left\{ (1-z)^{1-\beta/2} \right.
\\ \nonumber
&-& \frac{\beta(\beta-1) + 2h}{4}
\frac{\Gamma(1-\beta)}{\Gamma(2-\beta)} (1-z)^{2-\beta/2} 
\\ \nonumber
&+& \left. 
\frac{2^{-\beta} \pi \csc \beta \pi \Gamma(1+h+\beta)}
{\Gamma(1+h-\beta) \Gamma(1+\beta)}
 (1-z)^{1+\beta/2} + \cdots \right\} \; .
\eea
It can be noticed also in this limit that the two modes are
degenerate up to second order in $(1-z)$. 

The conclusion we can
draw from the solutions above at both asymptotic limits is 
that, in order to obtain a
pure mode at $t\rightarrow \infty$ one would need to fine-tune
the initial conditions to order $\exp [- 4(2-\beta/2)h|t|]$, and
inspect the final solution up to the same order and precision.
It should be evident that a numerical calculation would need
an astonishing level of accuracy to be able to detect such minute
differences. This explains why we could not address the question
of mode mixing in the numerical analysis.

With these solutions at hand, one can ask the question: is it actually
possible to avoid mixing when making a transition between two
contracting or expanding de Sitter phases? We see on inspection of
Eqs.~(\ref{Phi_1_minus}) to (\ref{Phi_2_plus}), expanding the
functions $1\pm z$ in time, that the general solution for the
gravitational potential reads
\be
\Phi^{(\pm)} \simeq A^{(\pm)} \ex^{\alpha_\pm t} + B^{(\pm)}
\ex^{\alpha_\pm t},
\label{Phi_pm}
\ee
where the coefficients $A^{(\pm)}$, $A^{(\pm)}$, and $\alpha_\pm$ can
be obtained formally from Eqs.~(\ref{Phi_1_minus}) to
(\ref{Phi_2_plus}). In the absence of mode mixing, one would expect
the transition matrix relating $\left\{ A^{(+)}, B^{(+)}\right\}$ to
$\left\{ A^{(-)}, B^{(-)}\right\}$ to be diagonal. As the coefficients
of the modes are different from one side to the other, this is clearly
not the case, so one expects mixing, at least in this simplified
model.

As a result, if the contracting and expanding phases are of comparable
durations, an initial condition in the contraction era which is a
mixed state of the dominant and subdominant modes will become a very
different mix of the dominant and subdominant modes at the end of the
contraction era, in effect transferring power from one component to
the other.  This transfer can result in a large amplification of
$\Phi$. This would mean that, provided there is a
scale-invariant spectrum in $\Phi$ before the bounce, however small its
amplitude, and even if it is present only in the sub-dominant mode,
the bounce can manage to amplify it to large values. However, this
mechanism does not separate the spectrum from the amplitude,
as, say, the curvaton models \cite{curvaton}, 
because the curvature perturbation is,
on large, cosmologically relevant scales, conserved through 
the bounce, and thus retains 
its amplitude as well as its spectral index.

\section{Spectrum of perturbations in a K-bounce}

There are two ways in which we can address the question about
cosmological perturbations in K-bounce models.  First, we could assume
that there were no perturbations initially, and that a spectrum of
cosmological perturbations was generated by the bounce itself, through
the usual quantum mechanism.  Second, we could equally well assume
that the bounce only distorts a pre-existing spectrum of cosmological
perturbations. Of course, in general both processes will occur, but in
linear theory they can be treated separately and the final spectrum
will be a combination of the two spectra. Let us briefly discuss the
first possibility of producing the perturbations at the bounce itself.

While tempting to produce perturbations close to the bounce, one
immediately faces a major difficulty, namely that it seems rather
unlikely that natural, vacuum-like, initial conditions could be
imposed close to the bounce.
Indeed, with the expansion
(\ref{H_bounce}), one has $z= a \sqrt{-2\dot H/H^2}$ in (\ref{v}), so
that switching back to the time variable $t$, we obtain
\be
\frac{z''}{z} = a^2 \left( \frac{\ddot z}{z} + H\frac{\dot z}{z} 
\right) \sim  a_0^2 \left( \frac{2}{t^2} - H_1 + \frac{H_2^2}{4
H_1^2} - \frac{H_3}{6 H_1} + \cdots \right),
\label{V_v}
\ee
where we have set, for simplicity, $c_X^2$ to unity, since we have
seen above (numerically) that this quantity is completely regular
through the bounce and actually hardly varies at all. To leading order
in $t$, Eq.~(\ref{v}) becomes
\be
\label{v_an}
\ddot v + H_1 t \dot v + \left( \frac{k^2}{a_0^2} -
\frac{2}{t^2}\right) v =0,
\ee
whose general solution is expressible in terms of hypergeometric
functions. For all but the smallest wavelength modes,
these happen to have no oscillatory part to which one
could connect the vacuum initial condition $v_\mathrm{ini}
=\ex^{-ik\eta}/\sqrt{2k}$. The evolution through the bounce itself is
therefore completely arbitrary for all scales of cosmological 
interest. Hence, in what follows we shall be
concerned exclusively with the second possibility, namely, 
of modifying a spectrum originally produced during the 
contracting phase prior to the bounce.

\begin{figure}
\center
\includegraphics[width=12.5cm]{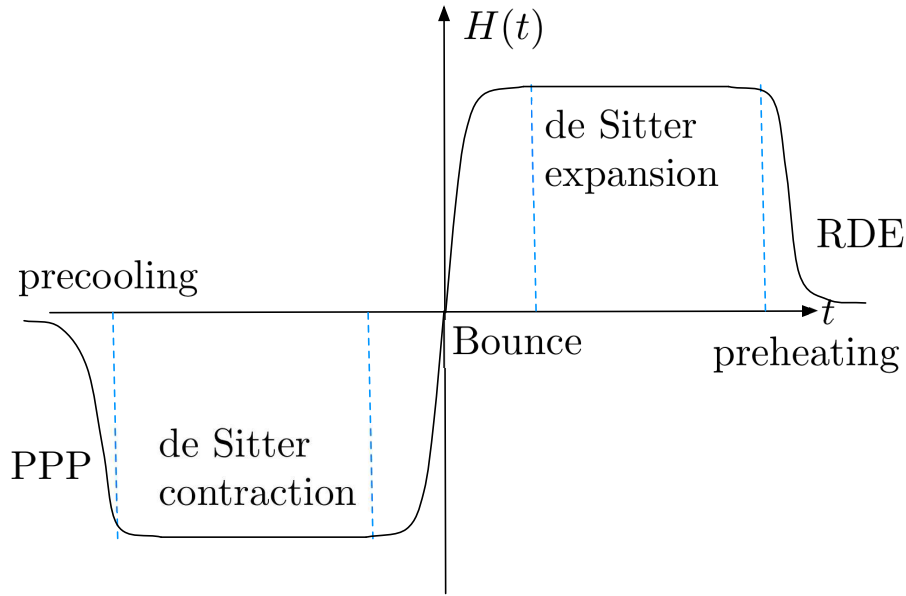}
\caption{\label{fig:4}
Typical embedding of the K-bounce into a complete cosmological
model. To begin with, the Universe is very large, empty, and
contracting. At that time, the cosmologically relevant modes are all
below their potential, \ie in practise their wavelength is smaller
than the Hubble scale, and a primordial spectrum of perturbation is
produced (region marked ``PPP'', standing for ``Primordial Perturbation
Production''). Then comes the precooling, when the initial era
condensates into the effective
non-canonical field $\phi$, which then starts to describe the 
cosmological dynamics. A contracting de Sitter stage follows, 
then the bounce and after that the Universe expands in a de Sitter
phase. This (presumably short) de Sitter era ends through 
a preheating mechanism, giving way to the
radiation-dominated era (RDE). 
}
\end{figure}

Recently, \cite{Creminelli} considered a K-essence model in which the
contracting phase prior to the bounce was described by $H \simeq p/t$
near the bounce, which leads to $n_{_\mathrm{S}}-1\simeq 2p$.
However, that approximation clearly breaks down at the bounce.  We
will now assume, in the same fashion, that a contracting phase not
described by our K-bounce model has already taken place, and that 
a radiation-dominated era starts shortly after the bounce.
The situation is summarized on Fig.~\ref{fig:4}.
We will simply assume that the curvature perturbations
are transmitted in a non-singular way through these precooling
and preheating phases -- as happens in the usual mechanism of
preheating. Therefore, here we
are just interested in the way an initial spectrum created in a
pre-bounce era (before precooling) is affected by the bounce,
and how that spectrum is transmitted to the radiation-dominated 
era after preheating. Given that all modes of interest are in
the infrared limit (long wavelengths) before and after the bounce,
we only need to ask what spectral distortions (if any) are
introduced by the bounce.

The overall conclusion we can draw from the last section is that an
initial spectrum of long-wavelength modes is entirely unaffected by
the bounce: the long-wavelength modes all behave in exactly the same
way, so their relative amplitudes remain unchanged. The curvature
perturbation $\zeta$ is conserved across the bounce for large enough
wavelengths. What emerges from the numerical analysis discussed in the
previous section is that for long, \ie cosmologically relevant,
wavelengths, the spectrum produced before the bounce, during the
contracting phase, is essentially unchanged apart from an overall
amplification factor
for $\Phi$ which, however, still leaves 
the curvature perturbation constant.
Let us consider for instance the case of
Fig. \ref{fig:2}, very close to the bounce, and suppose the
contracting phase lasted much longer in the past than indicated. Since
we are working with arbitrary units in time, we are free to assume
that the preheating-like phase begins, in this model, around $t\sim
7$, \ie very shortly after the bounce. Note that we must assume that
the expanding era is sufficiently close to the bounce in order to
ensure a natural solution to the flatness problem.

Connecting the bounce with the radiation-dominated phase, in a way yet
to elucidate, very shortly after this bounce took place, we end up
with a spectrum of perturbations which is undistorted.
(recall
that our initial conditions were such that $\Phi_\mathrm{ini}=1$ and
$\dot\Phi_\mathrm{ini}=0$.)  In other words, the transfer
matrix~\cite{courb1}, \ie the matrix that relates the initial
amplitudes of the growing and decaying modes to their final amplitudes
at some fiducial instant of time, does not depend on scale.  As we
have evolved the perturbations numerically, it is extremely difficult
to extract information about the decaying mode. Instead, we focus on
the spectrum of the perturbations at some point after the bounce has
occurred.

\begin{figure}
\center
\includegraphics[width=12.5cm]{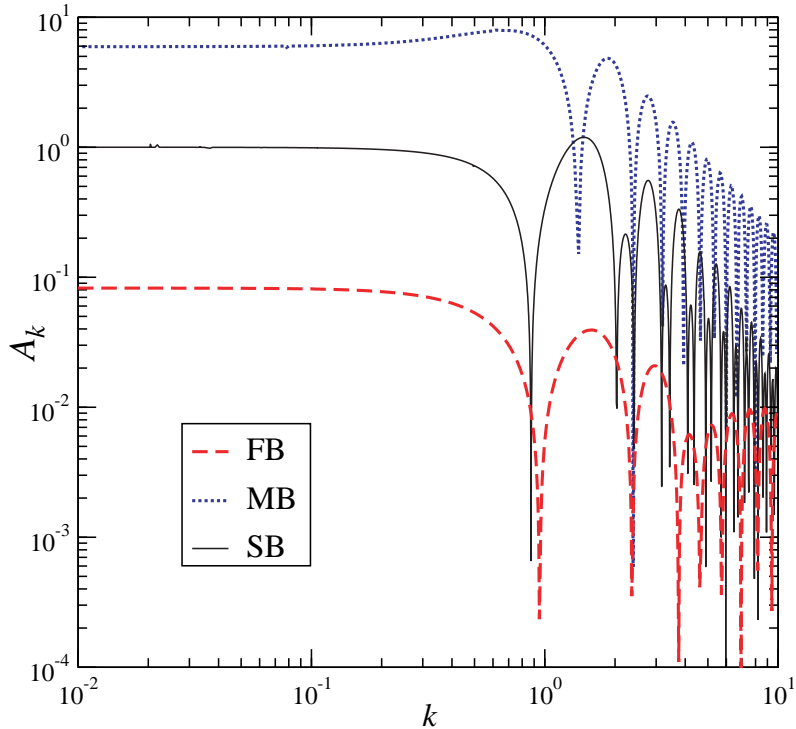}
\caption{\label{fig:5}
Spectral distortions for the FB, MB and SB models.  The onset of the
oscillatory regime is roughly at $k \sim 0.8 $ for the FB and
SB models, whereas for the MB model it lies at $k\sim 1.5$. The overall
normalization in this figure is arbitrary -- the curves only express
a change in the relative amplitudes of the modes.}
\end{figure}

There are two ways in which we can compute the spectrum. 
First, we can use the dominant solution of Eq. (\ref{Phi_SR}), 
and calculate its amplitude as a function of the
wavenumber $k$. By setting all modes to the same initial value, we
thus obtain the spectral distortions $A_k$ caused by the bounce. This
is shown in Fig. \ref{fig:5} for the concrete models we considered. 
The main result is that for long-wavelengths (small $k$) the
amplitudes are completely flat, meaning that the bounce does not
distort the initial spectra.
Notice also that, for the small wavelengths, 
the onset of the oscillatory regime influences
their relative amplitudes, and the change in their spectrum seems to
be model-dependent. In general, slow bounces seem to produce a
decay in the spectrum for small wavelengths, whereas fast bounces 
have little to no overall effect over the UV sector of the 
spectrum apart from some oscillations.

The second method is to evaluate the amplitude of the long-wavelength
modes after the bounce. This is useful in order to compute the
amplification factor $\Delta$ -- which is only meaningful for those
long wavelengths. We have obtained, for the three models we studied,
$\Delta_{\mathrm{FB}} \simeq 3\times 10^4$, $\Delta_{\mathrm{MB}}
\simeq 3\times 10^6$ and $\Delta_{\mathrm{SB}} \simeq 10^4$ for the
FB, MB and SB models respectively -- see Figs. 1-3. Note that these
values are very much dependent on the initial time we put the initial
conditions on. This means that in practice, much larger amplifications
could easily be achieved.
However, notice that the physically relevant function is the
curvature perturbation $\zeta$. We have found,
in agreement with \cite{courb2}, that $\zeta$ in fact
remains essentially constant -- even if $\Phi$ can be vastly 
amplified. This means that the physical
observables are unaffected by the bounce.

\section{Discussions and conclusions}

Bouncing models have been proposed as possible alternatives to
inflation. Even though such models seem to be able to solve at 
least some cosmological puzzles such as the horizon and flatness 
problems, many of them
still face a basic difficulty of producing an almost scale-invariant
spectrum of perturbation. The main reason for this failure is that there is no
generally agreed upon way of making a bounce. In particular, this
stems from the fact that General Relativity forbids a bounce to take
place without spatial curvature or violations of the
energy conditions.

Bouncing models have been built based either on a positive spatial
curvature~\cite{courb1,courb2} or many fluids~\cite{2fluids} -- one of them
having negative energy. The results that have been obtained up to now
are very strongly model-dependent, ranging from no effect whatsoever
(the modes passing unaltered through the bounce), to a complete
modification of the spectrum involving $k-$mode mixing. It has been
argued that both the mode mixing and/or the spectral modifications
could be due to either the spatial curvature and/or the presence of
many degrees of freedom, and hence of entropy
perturbations. Therefore, there has been no agreement about whether
bounces would lead inevitably to spectral distortions. Hence the need
to try and find a simple bouncing model with only one degree of
freedom and no spatial curvature in the framework of GR.

We have achieved the construction of one degree of freedom simple
bouncing models by means of a generalized scalar field theory. Our
models violate most energy conditions, but still we have $\rho \geq
0$. However, because the phantom barrier ($\rho+p<0$) cannot be
bypassed without some sort of singularity, such models cannot be very
realistic, and must be embedded into a more complete theory containing
at least the usual expanding radiation-dominated era.  As we have
shown, all our cosmological models flow to asymptotic de Sitter
solutions with $H>0$. This implies that the connection with the
radiation era must be realized through some sort of preheating
mechanism.  Similarly, the contracting de Sitter solution is a
repulsor from which all contracting solutions flow. This means that,
going backwards in time, all pre-bounce solutions must have initiated
from a contracting de Sitter stage. Again, in order to relate the
bounce to a contracting universe dominated by a regular fluid such as
dust or radiation, one must have a mechanism similar to preheating,
but going the other way around, that we have called precooling -- see
Fig. \ref{fig:4}.

We have found that, in these simple models, the propagation of
perturbations is highly non-trivial: although the transition matrix 
which relates the growing and decaying 
modes before and after the bounce is wavelength-independent
(no $k-$mode mixing in the terminology of the first of
Refs.~\cite{courb1}), it is however non diagonal, so there is
in general some amount of mixing between the two modes. 
Analytical calculations show that even an exponentially small
initial contribution of the sub-dominant mode can lead to a high
degree of mode mixing in the final spectrum of perturbations.
Hence, if the two modes have different spectral tilts in the
contraction era, the resulting spectrum will almost surely consist
of a superposition of the two spectra. 

We have also found that this mixing can lead to 
an amplification of the metric perturbations $\Phi$: any initial suppression of
the sub-dominant mode deep in the contraction era would be 
offset by an equal amount of growth prior to the bounce, 
leading to potentially large amplification factors 
for $\Phi$. However,
this does not impact the physically relevant curvature perturbation
$\zeta$, which remains essentially constant despite the growth of
$\Phi$, implying that the bounce does not affect physical observables.

\section{Acknowledgments}

R.A. would like to thank the Institut d'Astrophysique
de Paris, and P.P., the Instituto de F\'{\i}sica (Universidade de S\~ao
Paulo), for their warm hospitality. We very gratefully acknowledge
various enlightening conversations with J\'er\^ome Martin, Nelson
Pinto-Neto and Fabio Finelli. R.A. would like to thank Jo\~ao
C. A. Barata for key comments on the theory of confluent
hypergeometric functions. We also would like to thank CNPq, CAPES and
FAPESP (Brazil), as well as COFECUB (France), for financial support.

\end{document}